%Paper: gr-qc/9406040
%From: Madhavan Varadarajan <madhavan@phys.psu.edu>
%Date: Wed, 22 Jun 94 15:51:49 EDT

%Final version for bulletin board, Jun 22 1994

\magnification=1200
\centerline{\bf A Striking Property of the Gravitational Hamiltonian}
\bigskip
\bigskip
\centerline{Abhay Ashtekar${}^{1,2}$ and Madhavan Varadarajan${}^{1}$}
\centerline{\it${}^{1}\!$Center for Gravitational Physics and Geometry}
\centerline{\it Physics Department, Penn State, University Park,
PA 16802-6300}
\medskip
\centerline{\it ${}^{2}\!$Isaac Newton Institute for Mathematical Sciences,}
\centerline{\it University of Cambridge, Cambridge CB3 0EH}
\bigskip

{\sl A Hamiltonian framework for 2+1 dimensional gravity coupled with
matter (satisfying positive energy conditions) is considered in
the asymptotically flat context. It is shown that the total energy of
the system is non-negative, vanishing if and only if space-time is
(globally) Minkowskian. Furthermore, contrary to one's experience with
usual field theories, the Hamiltonian is} {\rm bounded from above}.
This is a genuinely non-perturbative result.

{\sl In the presence of a space-like Killing field, 3+1 dimensional
vacuum general relativity is equivalent to 2+1 dimensional general
relativity coupled to certain matter fields. Therefore, our expression
provides, in particular, a formula for energy per-unit length (along
the symmetry direction) of gravitational waves with a space-like symmetry
in 3+1 dimensions. A special case is that of cylindrical waves which
have two hypersurface orthogonal, space-like Killing fields. In this case,
our expression is related to the ``c-energy'' in a non-polynomial
fashion. While in the weak field limit, the two agree, in the strong field
regime they differ significantly. By construction, our expression yields
the generator of the time-translation in the full theory, and therefore
represents the physical energy in the gravitational field.}
\footnote{$^1$}{This
is a  detailed account of the results presented in the
Brill-Misner symposium at the University of Maryland in May 1993.}
\bigskip
\bigskip
{\bf 1. INTRODUCTION}

Spaces of solutions to Einstein's equations admitting isometries have
provided a useful and simplified arena to analyse a number of issues in
3+1-dimensional
classical and quantum gravity. An outstanding example is that of stationary
space-times where the presence of the time-like Killing vector field can be
used to introduce the notion of multipole moments, energy and angular
momentum being only the first of a doubly infinite series.  These moments
can then be used to characterize the space-time geometry completely [1].
This class of space-times is of particular interest in astrophysics
where sources can often be idealized as being stationary. Another
class of models of interest is provided by Bianchi cosmologies. In
this case, the isometries are space-like and Einstein's equations are
reduced to {\it ordinary} differential equations in time.  The
analysis  therefore simplifies considerably. In
some cases, the equations can be integrated completely by making
appropriate canonical transformations in the phase space and the
models can then be used to get insight into the conceptual issues of
non-perturbative quantum gravity such as the problem of time (see, e.g.
[2]). In other cases --particularly the Bianchi IX models--
the equations continue to be sufficiently complicated so as to provide
interesting examples of chaotic dynamical systems (see, e.g.[3]).
These analyses have shed considerable light on the ``generic
behavior'' of solutions of Einstein's equations as one approaches a
singularity. Neither of these truncations of full general relativity
is, however, well-suited to tackle dynamical issues that are related
to the fact that the gravitational field has an infinite number of
degrees of freedom: in the stationary context, there is no time
evolution while in the Bianchi models, the truncation is
so severe that one is left with only a finite number of degrees of
freedom.

There are two sets of dynamical issues that hinge on the presence of
an infinite number of degrees of freedom. The first refers just to the
classical theory: one would like to get insight into the nature of
gravitational waves beyond the linear approximation. Of particular
interest is the notion of energy in these waves and its properties.
The second is quantum mechanical: one would like to learn more about
the field theoretic difficulties associated with the existence of
infinitely many modes that can be excited. To study these problems, a
``midi-superspace'' of solutions to Einstein's equations was studied
in some detail in the seventies. It consists of solutions to
4-dimensional vacuum equations with cylindrical symmetry. In
this case, the field equations again simplify. Because there still
remain an infinite number of degrees of freedom, one continues to deal
with {\it partial} differential equations. However, the presence of
the two hypersurface orthogonal
Killing fields reduces the problem to that of solving the {\it
linear} wave equation for a scalar field  in a 3-dimensional
(fictitious) Minkowski space.  More precisely, given a solution to the
wave equation on a 3-dimensional Minkowski space, one can simply write
down a 4-metric with cylindrical symmetry which satisfies the full,
non-linear vacuum equations. In the classical theory, one is then
led to the issue of the physical interpretation.
What, in particular, is the energy carried by these waves? Is it
always positive?  Can one write down a simple ``mass-loss'' formula at
null infinity?

These and related problems were examined by several authors, in
particular, by Thorne [4] (using Cauchy surfaces), Stachel [5] (in
terms of fall-off at null infinity) and Kucha\v r [6] ( in the context of
canonical quantization).  The analysis is far from being
straightforward due to the following complication: since the solutions
have a ``translational Killing field'' ($\partial/\partial z$, parallel
to the axis of rotation), the solution can not be asymptotically flat
either at spatial or null infinity in 4-dimensions, whence the
standard machinery of the ADM framework (see, e.g., [7]) at spatial
infinity or of the Bondi-Penrose [8,9] framework at null infinity is
simply not available. Indeed, the symmetry considerations tell us that the
total energy in the wave must be infinite (or identically zero). The
physically meaningful quantity would be the energy per unit length along
$\partial /\partial z$. Thorne succeeded in making this notion --which
he called the ``c-energy'' --precise. His final expression can be
understood as follows.  Since the
4-metric is completely characterized by the solution to the wave
equation in a 3-dimensional, reference Minkowski space, one can
just compute the conserved energy of the scalar field (in Minkowski
space) and declare that to be the total energy of the cylindrical wave
per unit length. The quantity is then manifestly positive and, at
least intuitively, satisfies the anticipated mass-loss formula.
Furthermore, in the weak field limit, it reduces to the expected
expression. Kucha\v r's strategy [6] to the problem of quantization
can be understood in a similar fashion. The 4-metric can be gauge
fixed in such a way that the only degree of freedom in it is the
solution to the wave equation. Since one knows how to quantize the
free scalar field in 3-dimensional Minkowski space, one can take over
that operator-valued distribution $\hat\Phi$ and insert it in the gauge
fixed metric to provide the quantum operator corresponding to the
4-geometry. (The same strategy has been used [10] in the ``one
polarization Gowdy models'' which again have two commuting Killing
fields defined, where however, the spatial topology is that of a
3-torus rather than $R^3$. The fact
that the space-like sections are compact does give rise to  the usual
problems in the definition of total energy. However, the essence of the
quantization procedure is not affected.)

There is however another --and much more general-- strategy. It is
well-known (see, e.g., [11]) that in presence of a space-like Killing field,
Einstein's vacuum equations in 3+1-dimensions are equivalent to
Einstein's equations in 2+1-dimensions with a source consisting of a
triplet of scalar fields constituting a (SO(2,1)) non-linear sigma
model. If the 3+1-Killing field is ``translational'', it is natural to
expect the 2+1-dimensional fields to be asymptotically flat (both at
spatial and null infinity). For such fields, one can imagine extending
the 3+1-dimensional ADM strategy [7] to define conserved quantities at
spatial infinity and the Bondi-Penrose strategy [8,9] to define fluxes
of energy-momentum at null infinity. From the 2+1-dimensional
perspective, the Hamiltonian generating asymptotic time-translations
at spatial infinity would represent the total energy of the given
isolated system. From the 3+1-dimensional perspective, this would represent
the energy per unit length (along the Killing trajectories). There is a
similar dual interpretation of quantities at null infinity. Thus, one
can go back and forth between the two pictures. The strategy is
attractive becasue it avoids the introduction and use of fictitious
Minkowski spaces ---the energy would arise as the generator of time
translation directly in the {\it physical} picture.  It is also
natural from the perspective of quantum theory. Indeed, there exists a
non-perturbative quantization of 2+1 gravity without sources. The next
step from the 2+1-dimensional perspective is to bring in matter. An
exact Hamiltonian framework in the classical theory would be the first
step towards such an analysis.  Is the energy positive? If so, one can
hope that there would be no problems with stability and/or unitarity
of the quantum theory. Are the utraviolet difficulties different, now
that we have effectively a theory only in 2+1 dimensions?

Finally, this framework will encompass the cylindrical waves of [4,6]
as a special
case. These 3+1-dimensional space-times have {\it two} space-like
Killing fields which, moreover, are hypersurface orthogonal. In the
2+1 reduction with respect to the translational Killing field
($\partial /\partial z$), hypersurface orthogonality introduces a key
simplification: the triplet of matter fields reduce simply to a single scalar
field $\Phi$ (the log of the norm of the translational Killing field)
satisfying the wave equation in the (curved) 2+1-dimensional geometry.
The presence of the second Killing field then imposes a further
rotational symmetry, which in turn implies that the scalar field $\Phi$
satisfies the wave equation with respect to the physical 2+1-dimensional
geometry if and only if it does so with respect to a fictituous Minkowsian
metric on the 2+1-dimensional manifold; we can thus recover the
description used in the analysis of cylindrical waves. One can
therefore ask: Does the notion of energy obtained here from the
perspective of 2+1-dimensional gravity reduce to the c-energy? There
is no apriori reason why the two should be the same: while the
c-energy is the generator of time translations for the scalar field
propagating in a 2+1-dimensional Minkowski space, the new energy would
be the generator of time translations for the full 2+1 theory
consisting of gravity plus matter. Similarly, in the quantum theory,
new avenues open up.  From the work of Allen [12] we can deduce that
there {\it is} a consistent quantization of axi-symmetric 2+1 gravity
coupled to a scalar field satisfying the wave equation, which is
equivalent to Kucha\v r's [6] quantization of cylindrical waves. However,
{} from the general perspective of 2+1-dimensional gravity a number of
new avenues also become available.

These considerations suggest that it is natural to investigate a more
general ``midi-superspace'' consisting of 2+1 gravity
coupled to physically reasonable matter fields. The purpose of this
paper is to present a Hamiltonian framework for this system. In
particular, we will show that the Hamiltonian has two interesting
properties, one expected on physical grounds and the second somewhat
unexpected and, at least at first, quite surprising. The first
property is that, provided the matter fields satisfy a local energy
condition --which they do if they are obtained by a symmetry reduction
of the 3+1 theory-- the total energy is non-negative and vanishes if
and only if all matter fields vanish and space-time is globally
Minkwoskian. The second property is that the total energy is bounded
{\it from above} as well. More precisely, the canonical framework
breaks down beyond this limit in the sense that there is no function
which can generate ``the asymptotic time translation'' in the part of
the phase space where the bound would have been
violated. Thus, the Hamiltonian we obtain is quite different from the
c-energy. We will see that the two are non-polynomially related. In the
weak field limit, they agree. However, in the strong field limit,
there is quite a difference. The existence of an upper bound for the
Hamiltonian suggests that in the quantum theory, the ultra-violet
behavior may well be quite different from the one encountered in
the 3+1 theory. This difference may well be the underlying reason
behind the finding [13] that 2+1 gravity coupled to scalar fields
is perturbatively finite.

Can we intuitively understand the existence of this upper bound? We will
see that from a space-time point of view, when the bound on the
Hamiltonian is
violated, all resemblance to asymptotic flatness is lost in the sense
that the points ``at spatial infinity'' can be reached by curves of
finite length from any point in the interior.  The qualitative picture
of this ``closing up'' of space is perhaps not surprising in the light
of the work by Deser, Jackiw and 't Hooft [14] on 2+1 gravity in
presence of point particle sources. However, when one looks at the
issue in detail one sees that there are some important differences
between point particles and smooth field sources.  In geometrical
terms, the locations of point particles do not, strictly speaking,
belong to the space-time manifold since the geometry there has conical
singularities. As a result, a complete Hamiltonian description is
difficult to construct in that case: one must specify the boundary
conditions not only at spatial infinity but also at the location of
the point particles and then show that the resulting phase space has a
well-defined symplectic structure and continuous Hamiltonian flows.
Strictly speaking therefore, in the case of point particle sources, there
is no clearcut relation between the closing up of space and properties of
the Hamiltonian. On the other hand, when the sources are smooth fields,
the Cauchy surfaces can be taken to be topologically $R^2$ and geometry is
smooth everywhere.
There are no conical singularities in the physical space-time and no
points need to be excised.  Consequently, boundary conditions on all
fields need to be specified only at spatial infinity and the task of
constructing a complete Hamiltonian description is significantly
simpler. Finally, at first, the phenomenon of ``closing up'' of space
may seem [15] to have an analog also in the 3+1 theory in the vacuum
initial data studied by Brill [16]. However, in that case, space
closes up when the total energy is {\it infinite}, whence the result
is not surprising ---one does not expect a solution with infinite energy
to be asymptotically flat.  Here, the ``closing-up'' occurs for a {\it
finite} value of the total Hamiltonian and on general physical
grounds, without the knowledge of the asymptotic behavior of the
solutions to the constraint equations, it is hard to see why this
should occur. Indeed, for cylindrical waves, for example, one can
imagine just tuning up the value of the scalar field $\Phi$ by scaling
it by an arbitrarily large constant factor, thereby obtaining a
perfectly reasonable initial data for the scalar field (say of compact
support) with an arbitrarily large energy. It is therefore puzzling
at first that the total Hamiltonian can be bounded from above. What
actually happens
is rather subtle: the total energy is a {\it non-polynomial} function
of the c-energy --the Minkowskian energy of the scalar field $\Phi$--
and as one tunes the scalar field up, while the c-energy diverges, the
Hamiltonian tends to its upper bound.

In section 2, we will present the basic Hamiltonian framework. The
boundary conditions --and hence also the details of the construction--
are quite different from the 3+1 theory: there is no fixed, fiducial metric
to which all metrics approach at spatial infinity and extra care is needed
in a number of steps. In section 3, we show that the Hamiltonian is bounded
{} from above in the sense indicated above. In section 4, we establish
the positivity of the Hamiltonian. Again, while the general ideas are
similar to those used in the proofs in the 3+1 theory [17],
a number of subtle differences arise from the peculiar features of the
2+1- dimensional boundary conditions. In section 5, we summarize the main
results and discuss their implications to cylindrical waves.

This is a detailed account of the results presented by one of us (AA) at the
Brill-Misner symposium at the University of Maryland in May 1993.

\bigskip
\bigskip
{\bf 2. Hamiltonian framework for 2+1 gravity with matter fields}

Since we are interested in the Hamiltonian framework, we will assume
that space-times $M$ under consideration have a topology $\Sigma\times R$
where $\Sigma$ is an arbitrary but fixed non-compact 2-manifold, the
complement of a compact set of which is diffeomorphic to the
complement of a compact set in $R^2$. Thus, the topological
complications, if any, are confined to a world-tube in the
space-time $M$ with
compact spatial sections; outside this world-tube, $M$ resembles $R^3$.
The Cauchy surfaces in $M$ are to be diffeomorphic to $\Sigma$. In the
geometrodynamical framework, the basic phase space variables are the
2-metrics $q_{ab}$ on $\Sigma$ and their canonically conjugate
momenta, $P^{ab}$. The momenta are related to the extrinsic curvature
$K^{ab}$ via $P^{ab}= \sqrt{q}(K^{ab} -Kq^{ab})$, where $\sqrt{q}$ is the
square-root of the determinant of the metric $q_{ab}$, and $K=
q^{ab}K_{ab}$, the trace of the extrinsic curvature.

Our first task is to specify boundary conditions on $q_{ab}$ and
$P^{ab}$. As in 3+1-dimensional general relativity, the choice of
boundary conditions will be motivated by the asymptotic behavior of
simple exact solutions. Let us therefore make a small detour to
recall [14] the solution corresponding to a static point particle.

The spatial slice $\Sigma$ in this case is topologically $R^2$. Fix a
global ``Cartesian'' chart $x,y$ on $\Sigma$. The particle resides at
the origin, $x=y=0$. The solution to the Einstein's equation is given by:
$$ dS^{2}= -(dt)^{2}\;+\;r^{-8GM}[(dr)^{2}\;+\;r^{2}(d\theta)^{2}]
\eqno(2.1) $$
where $G$ is Newton's constant (which, in 3-dimensions has the
dimensions of inverse mass) and $(r,\theta)$ are the polar coordinates
obtained from the cartesian ones in the usual way (i.e.
$(x=r\cos\theta,y=r\sin\theta)$). Thus $r\;\epsilon \;[0,\infty)$ and
$\theta \;\epsilon \;[0,2\pi)$. A direct calculation shows that, as
required, the stress-energy tensor $T_{ab}$ is distributional and
localized at the origin; $T_{ab} = M\delta^2(x,y) \nabla_a t
\nabla_b t$.

Let us recast (2.1) in a more transparent form. Since the Ricci
tensor vanishes for $r>0$, and since we are in 3 space-time
dimensions, the full Riemann tensor vanishes there as well. Thus, for
$r>0$, the metric (2.1) is flat. To exhibit it in the manifestly flat
form, let us set
$$\alpha:=1-4GM, \quad {\rm and}\quad
\rho:={r^{\alpha}\over\alpha,} \;\;\;\; \bar{\theta}=\alpha \theta\ .
\eqno(2.2)$$
The metric (2.1) then takes the manifestly flat form:
%\begin{equation}
$$ dS^{2}=-(dt)^{2}\;+\; (d\rho)^{2}\;+\; (\rho)^{2}(d\bar{\theta})^{2},
\eqno(2.3)$$
where, however,  $0\leq \bar{\theta} \leq|\alpha|$. From the restricted
range of $\bar{\theta}$, we immediately see that there is a conical
singularity at $\rho=0$. We also see that the deficit angle is directly
related to the value of the mass $M$ of the particle. Thus, although the
space-time is flat, unless $M=0$ it is not {\it globally} isometric to the
3-dimensional Minkowski space. Indeed, the deficit angle persists even
at infinity. Thus, space-time metrics with different values of $M$
differ from each other already in the {\it leading order} terms at
infinity. In 4-dimensions, all asymptotically flat metrics approach a
fixed globally Minkowskian metric near infinity and the information
about the mass resides in the leading order, $1/r$-deviations from
this Minkowski metric. In 3-dimensions, by contrast, the information
about the mass resides already in the ``zeroth order'' behavior of the
metric at infinity; there is no universal, Minkowski metric that they
approach.

With these motivating remarks, we are ready to specify the boundary
conditions. Since, outside a compact set, $\Sigma$ is diffeomorphic to
the complement of a compact set in $R^2$, in the asymptotic region of
$\Sigma$ we can fix coordinates $r, \theta$, with $r_o<r<\infty$ and
$0\le \theta< 2\pi$. Let $x, y$ be the Cartesian coordinates
corresponding to $r,\theta$. Denote by $e_{ab}$ the Euclidean metric
defined by these coordinates; $e_{ab}= \nabla_a x\nabla_b x + \nabla_a
y \nabla_b y$. Note that the coordinates $(r,\theta)$, $(x,y)$ and the
metric $e_{ab}$ are defined only in the asymptotic region of
$\Sigma$ where there are no topological non-trivialities. We will
require that the metric $q_{ab}$ have the asymptotic form:
$$ q_{ab} = r^{-\beta}[e_{ab} + O(1/r)], \eqno(2.4)$$
for some real constant $\beta$ which we leave arbitrary for the time
being. (As in the 3+1-theory, the fall-off conditions will refer to
the components of the tensor field being considered in the Cartesian
chart $x,y$ which is fixed in the asymptotic region. Also, if $f\sim O(r)$,
we assume that the derivatives in the Cartesian chart fall off as:
$\partial_a f\sim O(r^{n-1}), \partial_a\partial_b f \sim O(r^{n-2})$, etc.)
Comparison with (2.1) leads one to expect that, in the final
picture, the ADM mass would be coded in $\beta$ through $\beta = 8GM$.
This will turn out to be the case. To begin with, we will allow $\beta$ to
assume negative values; it is the positive energy theorem of section 4 that
will force $\beta$ to be positive.

Thus, the gravitational or geometric part ${\cal C}^{\rm geo}$ of our
configuration space will consist of smooth
metrics on $\Sigma$ which have the asymptotic behavior given by (2.4).
Note that $q_{ab}$ is assumed to be smooth {\it everywhere} on
$\Sigma$; in particular, it can not have conical singularities such as
the one at the origin in the point particle geometry.  Put
differently, $\Sigma$ is assumed not have ``interior boundaries.''
This in particular means that we will only consider {\it smooth matter
sources}.  Had we allowed singular sources such as point particles,
the Hamiltonian analysis would have been significantly more difficult:
to obtain a consistent framework {\it all} sources must be included in
the construction of the phase space and it is generally difficult to do
symplectic geometry with singular fields. Thus, metric (2.1) was used
only to motivate the boundary conditions at {\it infinity}; in the
interior, the geometries included in the phase space will be quite
different.

Since the geometric part $\Gamma^{\rm geo}$ of the phase space is the
cotangent bundle over ${\cal C}^{\rm geo}$, it is completely determined
by ${\cal C}^{\rm geo}$. To exhibit the induced
boundary conditions on the momenta, let us first examine the
asymptotic behavior of tangent vectors $\delta {q}_{ab}$ at a generic
point $q_{ab}$ of ${\cal C}^{\rm geo}$. By varying (2.4) and using
$\ln r/r\sim 1/r$ for large $r$, we obtain:
$$ \delta q_{ab} \ \sim \  r^{-\beta}[ -\delta\beta ln (r) \ e_{ab}
+O(1/r)] \  .\eqno(2.5)$$
(Note that, unlike in the 3+1-theory, the fall-off of the tangent vectors
$\delta{q}_{ab}$ varies from point to point on ${\cal C}^{\rm geo}$.)
The momenta $P^{ab}$ are to be such that, regarded as cotangent
vectors, their action $P\circ[\delta q]$ on any tangent vector
$\delta q$ should be well-defined, i.e. the following integral should exist
$$P\circ[\delta q]:=\int_{\Sigma}\ d^2x \ P^{ab}\delta q_{ab}\ .$$
(Here and in what follows, we integrate  only scalar densities over
$\Sigma$; a fiducial volume element is therefore unnecessary.)
This requirement fixes the boundary conditions on $P^{ab}$:
$$P^{ab}e_{ab}\sim r^{\beta-3}, \quad P:=P^{ab}q_{ab}\sim r^{-3},
\quad {\rm and} \quad
[P^{ab}-\textstyle{1\over 2}P q^{ab}]\sim r^{\beta-2}\ .\eqno(2.6)$$
Thus, the phase space $\Gamma^{\rm geo}$ is to consist of smooth fields
$(q_{ab}, P^{ab})$, where $q_{ab}$ is a a positive definite metric on
$\Sigma$ and $P^{ab}$ a tensor density of weight one, satisfying the
boundary conditions (2.4) and (2.6). Since the momenta $P^{ab}$ have a
well-defined action on tangent vectors, it follows that the gravitational
part of the symplectic structure is well-defined.

Let us next consider matter fields. We do not wish to commit ourselves to
specific types of sources; our main restriction will only be that the matter
fields satisfy the following energy condition: $T_{ab}t^a n^b \ge 0$,
where $T_{ab}$ is the stress-energy tensor of matter and $t^a$ and $n^a$
are any future-directed time-like vector fields. Consequently, the form of
the boundary conditions will now be rather general. First, we require that
the fall-off on the fields and their momenta should be such that the matter
part of the symplectic structure is well-defined. Second, we demand
that the components of the matter stress-energy tensor $T_{ab}$ in
the Cartesian chart should be $O(r^{\beta -3})$. Note that it is easy to
satisfy this last condition ${\it and}$ have interesting solutions to the
constraints. For example, the matter
sources could have compact spatial support.
In this case, the space-time metric would be flat in a neighborhood of spatial
inifinity. Nonetheless --in contrast to the situation in the 4-dimensional case
where the constraints would have forced the initial data to correspond to
Minkowski space-time globally-- there is an infinite dimensional family of
non-trivial solutions to constraints (e.g. the ones corresponding to
cylindrical waves.)

To conclude this section, let us list some consequences of these
conditions which will be needed in the subsequent analysis. (2.4)
implies that the asymptotic behavior of $\sqrt{q}$ is given by:
$$\sqrt{q}\sim r^{-\beta}\ ,\eqno(2.7)$$
and that of the Ricci scalar is given by:
$$R\sim r^{\beta-3}\quad {\rm and}\quad \sqrt{q}R \sim r^{-3} .\eqno(2.8)$$
Finally, (2.6) implies that the asymptotic behavior of the extrinsic curvature
is given by:
$$ K_{ab} \sim r^{-2}\quad {\rm and} \quad K \equiv
         K_{ab} q^{ab}\sim r^{\beta -3}.
\eqno(2.9)$$

\bigskip\bigskip

{\bf 3. The constraints and the Hamiltonian}

Let us begin by recalling the situation in the 3+1-dimensional case in
the asymptotically flat context. The phase space has two sets of constraints,
a vector $C_a(x)$ and a scalar $C(x)$. To analyse the canonical transformations
they generate, one smears them by shift $N^a(x)$ and lapse $N(x)$ fields to
obtain functions $C_{\vec N}(q,p)$ and $C_{N}(q,p)$ on the phase space and
computes the corresponding Hamiltonian vector fields, i.e., infinitesimal
canonical transformations they generate. Now, because of the fall-off
conditions
(of the 3+1-dimensional theory) on the canonical variables, it follows that the
constraint functions fail to be differentiable unless the smearing fields $N$
and
$N^a$ go to zero at infinity. Thus, what constraints generate are spatial
diffeomorphisms and time-evolutions which are asymptotically identity. Assuming
these fall-offs on the lapse-shift pairs, one can compute the Poisson
brackets
between $C_{\vec N}$ and $C_{N}$. They constitute a first class system.
Hence the
canonical transformations they generate should be
thought of as gauge. The space-time
translations, on the other hand, correspond to lapse shift pairs
which are asymptotically constants. These do induce canonical transformations
on
the phase space but to obtain their generating functions, one must add suitable
boundary terms to  the smeared versions of the constraint functions.
Consequently,
even when the constraints are satisfied,the generating functions do not vanish;
they are simply reduced to surface terms, the ADM 3-momentum and energy [7].
To
summarize, on the mathematical side, there is a delicate interplay between the
boundary conditions and the differentiability of the constraint functions. This
in turn gives rise to a physical distinction between gauge and dynamics.
The former corresponds to spatial diffeomorphisms which are asymptotically
identity and the bubble-time evolutions which  fail to move the Cauchy surfaces
at infinity. The latter correspond to asymptotic space and time translations.
On
the constraint surface,
the numerical values of the generators of gauge transformations
vanish while those of dynamics are given by boundary terms.
Thus, there is a clean
separation between gauge and dynamics. (For further details, see, e.g. [19].)

The overall structure is similar in 2+1-dimensions. In section 3.1, we will
analyse the vector constraint and in 3.2, the scalar constraint. Section 3.3
discusses these results from various angles.
\bigskip

{\sl 3.1 The vector constraint}

Given a shift $N^a$ on $\Sigma$, the smeared vector constraint can be written
as:
$$ C_{\vec{N}}  = 2\int_{\Sigma} d^2x\  N^{a}D_{c}(P^{cd}q_{da})
+\ \hbox{\rm{matter terms}}\ . \eqno(3.1)$$
With our assumptions on the matter fields, the integral involving matter
fields is well-defined and will play no role in the discussion of this section.
We will therefore focus just on the gravitational part, i.e., the first term
on the right hand side of (3.1), which we will refer to as
$C_{\vec N}^{\rm geo}$. Using the boundary conditions (2.4) and (2.6) on
$q_{ab}$ and $P^{ab}$, we conclude that $D_b(P^{bc} q_{ac}) \sim O(1/r^3)$.
Since the volume element is $d^2 x = r dr d\theta$, it follows that
$C_{\vec N}^{\rm geo}$ is well-defined (i.e., finite) provided the shift
$N^a $ behaves asymptotically as $N^a\sim N^{a}_{o} (\theta) +0(1/r)$.
Thus, as far as the issue of existence of $C_{\vec N}^{\rm geo}$ is
concerned, we can let $N^a$ to be a vector field which remains
asymptotically bounded with respect to $e_{ab}$; it does not have to
vanish in the limit.

The question then is that of differentiability of $C_{\vec N}^{\rm geo}$.
Let us begin with differentiability with respect to $P^{ab}$.  Let us
first write (3.1) as an integral over the interior of the disc $r\le R_o$,
integrate by parts, use Stokes' theorem and then take the limit as
$R_o\to \infty$. We then have:
$$ C_{\vec N}^{\rm geo} = \lim_{R_0\to \infty}\big[- \int_{r\le R_o } d^2x \
({\cal L}_{\vec N}\  q_{ab})  P^{ab}
+2\oint_{r=R_o} d\theta r N^a P^{bc} q_{ab} \partial_c r \ ] .\eqno(3.2)$$
Now, the integrand in the surface integral behaves asymptotically as
$R_o^{-1}$; the surface integral therefore vanishes in the limit. Hence,
the expression of the smeared constraint reduces to:
$$ C_{\vec N}^{\rm geo} = -\int_\Sigma \ d^2x\  ({\cal L}_{\vec N}\ q_{ab})
P^{ab} .
\eqno(3.3)$$
The volume integral is now manifestly differentiable with respect to $P^{ab}$.
We have:
$${\delta C_{\vec N}^{\rm geo}\over{\delta P^{ab}}}= -{\cal{L}}_{\vec N}q_{ab},
\eqno(3.4)$$
which confirms our expectation that the canonical transformations generated
by $C_{\vec N}^{\rm geo}$ should correspond to the diffeomorphisms generated
by $N^a$ on $\Sigma$. Let us now consider differentiability with respect
to $q_{ab}$. For this, let us again write $ C_{\vec N}^{\rm geo}$ of
(3.3) as a limit of the integral over a disc $r\le R_o$, integrate by
parts and use Stokes' theorem to obtain:
$$  C_{\vec N}^{\rm geo} = \lim_{R_o\to \infty} \big[\int_{r\le R_o} d^2x\
q_{ab}
({\cal L}_{\vec N} P^{ab}) - \oint_{r=R_o} \ R_o d\theta N^c q_{ab}P^{ab}
\partial_c r \big]\ . \eqno(3.5)$$
Due to our boundary conditions, the integrand in the surface term now falls
off as $ R_o^{-2}$, whence in the limit, we have:
$$  C_{\vec N}^{\rm geo} = \int_\Sigma\  d^2x \ \ q_{ab}({\cal L}_{\vec N}
P^{ab}) \ . \eqno(3.6) $$
Thus, $ C_{\vec N}^{\rm geo}$ is now manifestly differentiable with respect to
$q_{ab}$:
$${\delta C_{\vec N}^{\rm geo}\over{\delta q_{ab}}}= {\cal{L}}_{\vec N}
P^{ab}\ , \eqno(3.7)$$
and the value of the derivative again confirms our expectation that
$ C_{\vec N}^{\rm geo}$ is the generator of spatial diffeomorphisms. For
simplicity, in the above discussion, we have left out matter terms. When
they are added, the total constraint functional $C_{\vec N} =
C_{\vec N}^{\rm geo} +C_{\vec N}^{\rm matter}$  generates diffeomorphisms
on the entire phase space consisting of geometrical and matter variables.

We conclude this subsection with two remarks.

1. While the general structure of the argument is the same as the one
normally used in 3+1 theories, there is nonetheless a key difference
in the final result. In the 3+1-analysis, the vector constraint generates
only those diffeomorphisms which are asymptotically {\it identity}.
These, in turn are interpreted as gauge. In the present case, we have
found that the vector constraints generates diffeomorphisms which can
remain {\it bounded} asymptotically; the shifts do not have to vanish
asymptotically. This conclusion may seem counter intuitive at first
since in the 3+1-theory, space translations on $\Sigma$ remain
asymptotically bounded and their generator on the phase space is the
ADM 3-momentum [7]. In the present case, on the other hand, due to the
presence of deficit angles at infinity, asymptotic space translations
are {\it not} symmetries of the theory. That is,unless $\beta =0$, the
translation Killing
fields of the fiducial $e_{ab}$ are not asymptotic Killing fields of
the $q_{ab}$ being considered because of the $r^{-\beta}$ term in the
boundary condition (2.4). The only asymptotic symmetries of the class
of space-times under consideration are the ones associated with time
translation and spatial rotation. (This observation was made by several
authors; see, in particular, [14,18]. Note incidentally  that, had we
introduced a $\theta$-dependence in the conformal factor relating
$q_{ab}$ and $e_{ab}$, we would not have had the rotational symmetry.)
Thus, there is complete consistency: there are neither space-translations
nor non-vanishing Hamiltonians associated with asymptotically bounded
diffeomorphisms which could have, potentially, played the role
of a (generalized) ADM 2-momentum.

2. We could have carried out the above analysis for a shift field $N^a$
which is an asymptotic rotational Killing field, i.e., behaves
asymptotically as $N^a \sim (\partial/\partial\theta)^a + O(1)$. We would
then have found that the surface term is non-zero, whence
$C_{\vec N}^{\rm geo}$ would not have been a differentiable function
on the phase space $\Gamma$. Thus, the asymptotic rotation is {\it not}
generated by the constraint; it does {\it not} correspond to a gauge
transformation. Indeed, it is easy to find the Hamiltonian $J_{\vec N}$
on the phase space $\Gamma$ which generates the corresponding canonical
transformation. As in the 3+1-theory, one just has to add to the contraint
functional the appropriate boundary term to restore differentiability
and rescale the result by $1/16\pi G$ to conform to the standard
normalization (which comes from the overall constant in the expression
of the action):
$$ \eqalign{J_{\vec N} &= \textstyle{1\over 16\pi G}[ \int_\Sigma\ d^2x\
q_{ab}\ ({\cal L}_{\vec N} P^{ab})   + \hbox{\rm matter terms}]\cr
   &\approx \textstyle{1\over 8\pi G} \oint dS_c N^a P^{cd}q_{da}\ , \cr}\
\eqno(3.8) $$
where $dS_c = r\partial_c r d\theta$ is the line element on the boundary
that arises in the Stokes' theorem and it is understood that the expression
is first evaluated on circle $r=R_o$ in the asymptotic region and then the
limit $R_o\to \infty$ is taken. The last step provides the numerical value
of the the angular momentum $J_{\vec N}$ {\it on the constraint surface}.
Our boundary conditions ensure that the integral is well-defined over
the entire phase space $\Gamma$. As in the 3+1 theory, the surface
integral involves only the gravitational variables; the matter terms enter
only through the constraints. The expression (3.7) agrees with the
formulas for angular momentum given by Deser, Jackiw and 't Hooft [14]
and by Henneaux [18].
\bigskip

{\sl The scalar constraint}

The steps in this analysis are the same as in the previous subsection.
However, since the final result is somewhat unexpected, we will provide
the relevant details.

Given a lapse function $N$ on $\Sigma$, we can write the smeared
constraint function as:
$$ \eqalign{C_N &= \int_{\Sigma} d^2x\  N [\sqrt{q}R -
{1\over\sqrt{q}}(P^{ab}P_{ab}
   - P^{2})]+ \hbox{\rm matter terms}\cr
   &= C_{N}^{\rm geo} + C_{N}^{\rm matter} \cr}\ .\eqno(3.9)$$
Again, matter terms will play no role in our discussion. Now, from (2.8)
we see that the ``potential term'' $\sqrt{q}R$ falls off as $r^{-3}$,
independently of the value of $\beta$, while (2.4), (2.6) and (2.7)
imply that the  fall-off of the ``kinetic term'' does depend on $\beta$:
$(1/\sqrt{q})(P^{ab}P_{ab} - P^2) \sim r^{\beta -4}$. Hence, the integral
containing the potential term will exist only if $N/r$ goes to zero,
while that containing the kinetic term will exist if
$Nr^{\beta-2}$ goes to zero, in the limit $r\to\infty$. Now, an
asymptotic time translation corresponds to $N\sim 1+ O(1/r)$. We therefore
see that {\it the kinetic integral will not exist for the $N$ corresponding
to time-translations unless $\beta$ is less than $2$}. Furthermore, the
addition of surface terms can not improve the situation since the kinetic
terms are purely algebraic in the canonical variables.
{}From the phase space viewpoint, this, in essence, will turn out to be
the reason why the Hamiltonian is forced to be bounded from above.

Next, we turn to the analysis of differentiability. It is straightforward
to verify that if we demand that the lapse go to zero asymptotically as
i) $O(1/r)$, if  $\beta\le 2$, and ii) $O(r^{-\beta +1})$ if $\beta\ge 2$,
not only does $C_{N}^{\rm geo}$ exist but it is also differentiable on
the phase space. The canonical transformations it generates correspond
to ``bubble time evolutions;'' the time translation vanishes identically
at infinity. As in the 3+1-dimensional theory, these correspond to
``gauge motions'' in the sense that they can be taken care of by appropriate
gauge fixing. (For details on this interpretation, see, e.g. [19]).

Let us now consider lapse functions which correspond to time translations,
i.e., have the asymptotic behavior $N\sim 1 +O(1/r)$. Using evolution
equations of the initial value formulation, we can formally write an
infinitesimal transformation on the phase space corresponding to this
time translation. Since
the evolution equation (for zero shift) on $q_{ab}$ is simply
$$ \dot{q}_{ab} = 2NK_{ab} \equiv \textstyle{1\over \sqrt{q}} (P_{ab} -
Pq_{ab})\ ,$$
it follows that, if a Hamiltonian is to exist, its kinetic piece must
be the same as that in (3.9). Since this does not converge for
$\beta\ge 2$, it follows that on the $\beta\ge 2$-part of the phase space,
there is simply no Hamiltonian which can generate a canonical
transformation corresponding to this evolution. Thus, while one can
formally write down the ``evolution equations,'' they do not induce canonical
transformations on this part of the phase space. We will discuss this
point in some detail at the end of this section.

{}From now on, therefore, let us focus on the ``physical'' part of the phase
space defined by $\beta<2$.

Now, if $N\sim N_\infty +O(1/r)$, where $N_\infty$ is a constant, the
functional $C_N^{\rm geo}$ does exist on the physical part of the phase
space. However, as in the 3+1-dimensional case, it is not differentiable.
Thus, again the ``evolution equations'' are not generated by the scalar
constraints. However, as in the 3+1-dimensional theory, this evolution
{\it does} correspond to a well-defined canonical transformation and its
generator is obtained by adding a suitable surface term to the constraint
functional. Let us now see how this arises. It is clear by inspection
that the kinetic integral, being algebraic in the canonical variables,
is differentiable with respect to both $q_{ab}$ and $P^{ab}$. Thus, we
can focus just on the potential term. Furthermore, since this term is
independent of momenta, we need only be concerned with its derivative with
respect to the configuration variable $q_{ab}$. Taking the variation of
the potential term, we obtain:
$$\eqalign{\delta \int_{\Sigma}\ d^2x N &\sqrt{q}R \ = \ \int_{\Sigma}d^2x \
\sqrt{q}(-D_{a}D_{b}N + D_{c}D^{c}Nq_{ab})\ \delta q^{ab}\cr
&+\oint_{r=\infty}\ d\theta \sqrt{h} [Nv_{a}+(D_{a}N)q^{bd}\delta
q_{bd}-D^{c}N\delta q_{ac}] r^{a}, \cr} \eqno(3.10)$$
where,
$$ v_{a}=D^{b}\delta q_{ab}-D_{a}(q^{bd}\delta q_{bd}), $$
$r^a$ is the unit normal to the circle at spatial infinity and $\sqrt{h}$
is the determinant of the induced metric, $h_{ab}$, on this circle.
(As before, it is understood that the integrals are first evaluated at a
fixed radius where integrations by parts are carried out and then the
radius is made to approach infinity. Also, in (3.10), we have used the fact
that the Einstein tensor $R_{ab} -\textstyle{1\over 2}R g_{ab}$ vanishes
identically in two dimensions.) The second and the third terms in
the surface integral --involving derivatives of the lapse function-- vanish
identically because of the choice of the boundary conditions while the
first term can be simplified. The final result is:
$$ \delta \int_{\Sigma} d^2x N \sqrt{q}R  =  \int_{\Sigma}\ d^2x
\sqrt{q}(-D_{a}D_{b}N + D_{c}D^{c}Nq_{ab})\ \delta q^{ab} +
\delta\beta \oint d\theta N_\infty \ .
\eqno(3.11)$$
It is the presence of the surface term that spoils the differentiability
of $C_{N}^{\rm geo}$. In the case when $N$ goes to zero, the surface
integral vanishes and $-C_{N}^{\rm geo}$ generates canonical
transformations corresponding to the ``bubble time evolution'' by an
amount dictated by $N$. Hence, we have an obvious strategy to obtain
the generator of the canonical transformation: substract the surface
integral from $-C_{N}^{\rm geo}$. This strategy does work and the
Hamiltonian generating the time translation which is unit at infinity
is given by:
$$H = -\textstyle{1\over 16\pi G} \big[C_N^{\rm geo} - C_N^{\rm matter} -
 \beta \oint_0^{2\pi}\ d\theta \ ]\ ,
 \eqno(3.12)$$
where, we have again divided by the factor $1/16\pi G$ to conform to
the standard normalization.

Let us summarize. In the part of the phase space corresponding to
$\beta \ge 2$, ``time translations'' do not induce canonical
transformations; there is no Hamiltonian generating them. In the part
with $\beta<2$, the Hamiltonian does exist and is given by (3.12). On
physical states, constraints are satisfied and its numerical value is
given simply by $\beta/8G$. Together, these results lead us to the
conclusion that the Hamiltonian is bounded from above; $H< 1/4G$.
\bigskip

{\sl Discussion}

1. The result that the Hamiltonian is bounded from above is quite
unsettling at first. Let us therefore probe it from various angles.

In certain Bianchi II models, although the space-time picture and the
initial value formulation are perfectly well defined,the standard
ADM type Hamiltonian formulation fails to exist (see, e.g., [20]) and,
as of now,
one does not have viable replacements. Is the situation similar here?
That is, does the main result of this subsection have to do only with
the Hamiltonian framework or does something strange happen at
$\beta =2 $ also from the viewpoint of space-time geometry? Let us begin
with the point particle example [14] discussed in the beginning of
section 2. For $\beta<2$, there is a conical singularity at the origin.
At $\beta=2$, on the other hand, the cone simply opens up to become a
cylinder and the distinction between the origin and infinity is blurred.
For $\beta>2$, the old origin becomes the point at infinity and the
particle can be thought of as residing at the old point at infinity.
Thus, something strange does happen to the geometry. However, in our case,
there are only smooth matter sources and, in particular, there are no
conical singularities or even preferred points on $\Sigma$. Therefore the
point particle picture can only be taken as an indication.

This indication is correct:
Something remarkable does happen to the spatial geometry at $\beta =2$
even in the smooth case. For $\beta <2$ the points $r=\infty$ are,
as one would expect, at an infinite proper distance from any point in
the interior. Indeed, assuming that the matter sources have compact
support, one can calculate the geodesic distance from any point in the
asymptotic region to a point at infinity and find that it diverges as
a power of $r$. For $\beta=2$, the divergence is logarithmic. For
$\beta >2$, there is no divergence; the points $r=\infty$ {\it are at
a finite distance with respect to any point on} $\Sigma$. Thus, for
$\beta >2$, space simply ``curls up'' and there is no resemblance to
asymptotic flatness. Note, however, that $\Sigma$ is {\it not}
compactified; the 2-metric $q_{ab}$ does not extend to ``the point
at infinity'' in a smooth manner. $\Sigma$ is still non compact but
it is geodesically incomplete; there is, effectively, a singularity
at ``the point at infinity.''

Indeed, it is not clear if there are any physically admissible solutions
to the constraints on the Cauchy surface $\Sigma$
when $\beta \ge2$. The simplest case would be to
consider matter fields in the 2+1-dimensional theory which arise from
the symmetry reduction of 3+1-dimensional cylindrically symmetric
space-times. In this case, global analysis has recently been
carried out, without requiring that the two Killing fields be
hypersurface orthogonal [21]. It was found that asymptotically flat
solutions to constraints exist only if $0\le \beta < 2$. This may seem
surprising at first since one might expect the energy to grow unboundedly
as one keeps scaling the matter fields by a constant. However, as
one ``tunes up'' the matter fields, due to the curling up of $\Sigma$,
the effective gravitational ``potential energy'' also goes up such that
$\beta$ remains bounded below $2$. (This point is discussed further
in section 5.)

Even if one assumes that physically reasonable solutions with $\beta>2$
exist to the constraint equations, at least on heuristic grounds it
would appear that, due to the effective singularity at the point at
infinity, difficulties should arise with finite evolution. Given any
$\epsilon >0$, one would expect under the evolution by proper time
$\epsilon$, singularities would appear in the neighborhood of the point
at infinity of radius $\epsilon$ since ``the past light cones of points
in this neighborhood would contain the singular point at infinity.''

Thus, it is reasonable to expect that the difficulties we encountered
at $\beta =2$ are not artifacts of the Hamiltonian formulation. These
points of the phase space are pathological also from the viewpoint of
space-time geometry.

2. Let us restrict oursevelves to the part of the phase space where
$\beta <2$. It is easy to check that the Hamiltonian $H$ and the angular
momentum $J$ have vanishing Poisson brackets with all the constraints.
Hence, they are gauge invariant; they are observables of the theory in
the sense of Dirac [22]. Since $H$ and $J$ are differentiable functions
on the (restricted) phase space, one can take their Poisson bracket.
It vanishes, reflecting the fact that the time translation and the rotation
symmetries commute. Thus, the overall picture is internally consistent
and the situation is completely analogous to that in the 3+1 theory.

3. The surface term which provides the numerical value of the Hamiltonian
on the constraint surface agrees with that of Deser, Jackiw and
't Hooft [14] and of Henneaux [18]. As was pointed out by Henneaux,
in 2-dimensions, the Ricci scalar is a pure divergence and therefore
can be expressed as a surface term which, apart from an overall constant,
coincides with the surface term in the Hamiltonian. Note however that
the Hamiltonian is {\it not} given by the integral of the Ricci scalar;
indeed, as we saw above this term is not even differentiable with respect
to $q_{ab}$. Thus, to obtain the correct evolution even at points of the
constraint surface, we must use the full Hamiltonian given in (3.12).
Finally, after this work was completed, it was pointed out to us that
a number of authors had noticed that in special contexts
--such as cylindrical symmetry [21], time symmetric initial data for
cosmic strings [23], etc -- the deficit angle at infinity is bounded both
{} from above and below. However, the generality of the result  and
especially its
relation to the boundedness of the Hamiltonian generating time translations
in a proper phase space formulation was not analysed in these references.
\bigskip\bigskip

{\bf 4. Positivity of energy}

In the previous section we saw that the Hamiltonian is bounded from above.
We now wish to show that it is also bounded below; when the constraints
are satisfied with matter fields satisfying our energy condition,
the Hamiltonian is non-negative and vanishes if and only if the matter
fields vanish and the initial data is that of Minkowski space. Under
certain restrictive assumptions, positivity was established by a number
of authors. For example, if one restricts oneself to matter fields that
arise from a symmetry reduction of 3+1-dimensional cylindrically symmetric
space-times, a theorem due to Berger, Cru\'sciel and Moncrief [21]
ensures that $\beta \ge 0$ and vanishes if and only if we are in
Minkowski space. Similarly, Henneaux [18] has observed that since the
surface integral in the expression of the Hamiltonian is proportional
to the integral $\int d^2x\sqrt{q}R $, it is straightforward to establish
positivity of the Hamiltonian on a $K=0$ surface. Here, we will use
technniques similar to those introduced by Witten [17] in the 3+1 theory
to establish positivity without such restrictions.

In the first part of this section, we recall basic facts about
$SU(1,1)$ spinors and in the second part, establish the main result.
\bigskip

{\sl 4.1 $SU(1,1)$ spinors}

Since the reader is likely to be more familiar with $SU(2)$ spinors than
$SU(1,1)$, we will adopt conventions that are geared to the $SU(2)$ case.

Let us begin my recalling the elements of spinor algebra. Let $S$
denote a 2-dimensional, complex vector space and let
$\alpha^A, \beta^D, ...$ denote its elements.
These will be called (one index) spinors. Let
us fix a second rank, non-zero tensor $\epsilon^{AB}$ over $S$ and denote
its inverse by $\epsilon_{AB}$; thus $\epsilon_{AC} \epsilon^{BC} =
\delta_A^B$. Following the Penrose-Rindler [24] convention, we will raise
and lower the spinor indices using these tensors: $\alpha^A\epsilon_{AB}
= \alpha_B$ and $\epsilon^{AB}\alpha_B = \alpha^A$. Next, we introduce a
Hermitian conjugation operation $\dagger$ on $S$ satisfying:
$$(\alpha^A + k \beta^A)^\dagger = (\alpha^\dagger)^A +
\bar{k}(\beta^\dagger)^A
\quad (\alpha^\dagger)^\dagger = - \alpha^A\quad {\rm and}\quad
\alpha^{\dagger A} \alpha_A \ge 0 \ , \eqno(4.1)$$
where, $k$ is any complex number and the equality in the last condition
holds if and only if $\alpha^A =0$. We then extend this operation to
spinors of arbitrary rank by demanding that
$$ (\epsilon^\dagger)_{AB} = \epsilon_{AB}, \quad {\rm and}\quad
(\alpha_{A...B}^{C...D}\ \beta_{M...N}^{P...Q})^\dagger =
\alpha^{\dagger}{}_{A...B}^{C...D}\ \
\beta^\dagger{}_{M...N}^{P...Q}\  .\eqno(4.2)$$
(For details, see, e.g., [25], chapter 5.)

We can now consider the space $V$ of trace-free Hermitian, second rank
spinors $\alpha_A{}^B$. $V$ is a 3-dimensional, real vector space,
equipped with a natural, positive definite inner product: $(\alpha,
\beta) := -\alpha_A{}^B \beta_B{}^A \equiv -{\rm tr}\ \alpha\beta$.
To define $SU(2)$ spinor fields on a 3-dimensional Riemannian manifold,
one sets up a metric preseving isomorphism between $V$ and the tangent
space at each point of the manifold. In our case, the space-time manifold
$M$ is equipped with a metric of signature $-++$ and hence a ``Wick
rotation'' is required. To accomplish this, choose a spinor $n_{AB}$
satisfying: $n^\dagger_{AB} = -n_{AB},\ n_A{}^A = 0, {\rm and}\ {\rm tr}
\ n.n =1$. Denote the one dimensional, real subspace
spanned by the real multiples of $n^{AB}$ by ${\cal N}$ and the
2-dimensional real subspace of {\it Hermitian} spinors $\alpha_A{}^B$
which is orthogonal to $n_{AB}$ (so ${\rm tr}\ \alpha n = 0$) by
${\cal V}$  and let $ V' = {\cal N} \oplus {\cal V}$.
Then, $V'$ is a real; 3-dimensional vector space, equipped with a natural
metric of signature -++ . To define $SU(1,1)$ spinor fields, one then
has to fix a metric preserving isomorphism --or, soldering form--
$e_a^{AB}$ between $V'$ and the tangent space at each point of $M$.

Since we are interested in the canonical framework, however, we will need a
slightly weaker structure. Let us fix a foliation of $M$ by space-like
2-manifolds $\Sigma$ and denote by $n^a$ the vector field which is unit,
future pointing and everywhere orthogonal to $\Sigma$. Let us assume
that $e_a{}^{AB}$ is so chosen that it maps $n^a$ to $n^{AB}$. Set
$E_a{}^{AB} = e_a{}^{AB} + n_a n^{AB}$ . Then, $E_{aA}{}^{B}$ solders
the 2-dimensional, real tangent space at any point of $\Sigma$ to the
vector space ${\cal V}$ of trace-free, Hermitian spinors which are
orthogonal to $n_{AB}$. In particular, therefore, the 2-metric $q_{ab}$
on $\Sigma$ is given by: $q_{ab} = -{\rm tr}\ E_aE_b$. For future use,
we note the following algebraic properties of the $\Sigma$-soldering
forms $E_{aA}^B$ which will be useful in the next sub-section:
$$\eqalign { E_{aA}{}^{C}E_{bC}{}^{B} &= -\textstyle{1\over 2} q_{ab}
\delta_{A}^{B}\
+ \ \textstyle{i\over\sqrt{2}} \epsilon_{ab}n_{A}{}^{B}\cr
E_{mA}{}^{C}n_{C}{}^{B} &= {i\over\sqrt{2}} \epsilon_{m}{}^a E_{aA}
{}^{B}\cr} \eqno(4.3)$$
where $\epsilon_{ab}$ is the alternating tensor on $\Sigma$ compatible
with $q_{ab}$. This completes the discussion of spinor algebra.

We can now introduce the basic notions of spinor calculus. First, it
is straightforward to establish that $\Sigma$ admits a unique derivative
operator $D$ (which acts on both the spinor and the tensor indices)
which is compatible with the given $E_{aA}{}^B$: the equation
$$0=D_aE_{bA}{}^B = \partial_a E_{bA}{}^{B} + [\Gamma_a, E_b]_A{}^B
-\Gamma_{ab}^c  E_{cA}{}^B  \eqno(4.5)$$
determines the Christoffel symbols $\Gamma_{ab}{}^c$ and the spin connection
$\Gamma_{aA}{}^B$ uniquely. A particularly convenient connection
$A_{aA}{}^B$
on spinors is obtained by adding to the spin connection a suitable multiple of
the extrinsic curvature $K_{aA}{}^B := K_{ab}E^b_A{}^B$:
$$ G A_{aA}{}^B := \Gamma_{aA}{}^B - {i\over\sqrt{2}}K_{aA}{}^B .\eqno(4.6)$$
In the 2+1-dimensional theory, $A_{aA}{}^B$ will play essentially the same role
as that played by the Sen connection in the 3+1-dimensional general relativity.
We will therefore refer to it as the Sen connection also in the present case.
Denote by ${\cal D}$ the derivative operator on spinors defined by the Sen
connection: ${\cal D}_a\alpha_A:= \partial_a \alpha_A + G A_{aA}{}^B\alpha_B$.
Then, we have the following two useful properties:
$$\eqalign{({\cal D}_a \alpha_A )^\dagger =: {\cal D}^\dagger_a
\alpha^\dagger{}_A & = {\cal D}_a\alpha^\dagger{}_A +i\sqrt{2}K_{aA}{}^B
\alpha^\dagger{}_B \cr
{\cal D}_aE^a_A{}^B &= 0\cr} ,\eqno(4.7)$$
where, in the second equation, we have used the fact that $K_{ab}$ is
symmetric and $K_{ab}n^b = 0$. Finally, as in the 3+1-dimensional theory,
the constraints can be expressed succinctly in terms of the curvature of
the Sen connection. We have:
$$ {\rm tr}\ E^aF_{ab}  = 4\sqrt{2}\pi i G T_{ab} n^a, \quad
{\rm and}\quad
{\rm tr}\ E^aE^bF_{ab} = 8\pi G T_{ab}n^a n^b .\eqno(4.8)$$
\bigskip

{\sl 4.2 Positivity of the Hamiltonian}

With the machinery of $SU(1,1)$ spinors at hand, we are now ready to
prove the positive energy theorem.  The beginning of the argument is
the same as in the 3+1 theory. However, due to the difference in the
boundary conditions, there is a departure from the 3+1-dimensional
procedure at an intermediate stage and, unlike in the 3+1-dimensional
case, the proof is now by contradiction.

Let us begin with the analog of the 3+1-dimensional Witten identity [17].
For any spinor field $\lambda_A$ on $\Sigma$, we have:
$$\eqalign{(E^b_D{}^A{\cal D}^\dagger{}_b) &(E^a_A{}^B {\cal D}_a )
\lambda_B = -\textstyle {1\over 2} {\cal D}^{\dagger a}\ {\cal D}_a
\lambda_D + E^a_D{}^A E^b_A{}^B {\cal D}_{[a}{\cal D}_{b]}\lambda_B \cr
&= -\textstyle {1\over 2} {\cal D}^{\dagger a} {\cal D}_a \lambda_D -
\textstyle{i\over{4\sqrt{2}}}\epsilon^{ab} F_{ab}{}^{EF}n_{EF}\lambda_D
+\textstyle{1\over 4} \epsilon^{ab}F_{ab}{}^{EF}E^m{}_{EF}
\epsilon_{mn}E^n_D{}^A \lambda_A\cr
&=-\textstyle {1\over 2}{\cal D}^{\dagger a} {\cal D}_a \lambda_D +
4\pi G (T_{ab}n^an^b \lambda_D - \textstyle{i\sqrt{2}}E^a_D{}^A
\lambda_A T_{ab}n^b) \ ,\cr}\eqno(4.9)$$
where, in the last step, we have used (4.3) and (4.8). Now, let us choose
for the spinor field $\lambda_D$, a solution to the analog of the
3+1-dimensional Witten equation:
$$E^a_A{}^B{\cal D}_a \lambda_B = 0 ,\eqno(4.10)$$
so that the left side of the Witten identity (4.9) vanishes. Then, if we
multiply both sides by $\lambda^{\dagger D}$, integrate over the disc
$r\le R_o$, and use Stokes' theorem to simplify the first term on the right
side, we obtain:
$$\eqalign{
\oint_{r=R_o}d\theta &\sqrt{h} r^{b}(\lambda^{\dagger D}{\cal D}_{b}
\lambda_{D})\cr
 &= \int_{r\le R_o} d^2x \sqrt{q} ({\cal D}_{b}\lambda^{D})^{\dagger}
({\cal D}^{b}\lambda_{D}) \
 +4\pi G\int_{r\le R_o} d^2x\sqrt{q} T_{ab}n^a(Nn^b+N^{b}),\cr}
\eqno(4.11)$$
where,
$$N = (\lambda^\dagger)^D \lambda_D\quad {\rm and}\quad N^a = i\sqrt{2}
E^{n\ AB} \lambda_A \lambda^\dagger_B. \eqno(4.12)$$
It is straightforward to verify that ($N^a$ is real and) $N^a+Nn^a$ is
time-like.

The idea now is to take the limit of this equation as $R_o\to\infty$.
For this, we need a control over the asymptotic behavior of the solution
$\lambda_A$ to (4.10). In the proof of the 3+1-dimensional positive energy
theorem, one simply requires $\lambda_A$ to asymptotically approach a
constant spinor. In the present case, however, a more subtle choice is
necessary because of the difference in the asymptotic conditions on the
metric and the extrinsic curvature. Fortunately, the analysis is
simplified because equation (4.10) is again conformally invariant (with,
however, a conformal weight for $\lambda_A$ which is different from the one
of the 3+1-dimensional theory):
\item{}{\sl $\lambda_A$ solves (4.10) with respect to $(q_{ab}, K_{aA}{}^B)$
if and only if $\phi^{-1/2}\ \lambda_A$ solves (4.10) with respect to
$(\phi^2 q_{ab}, K_{aA}{}^B)$ for any smooth, positive
function $\phi$ on $\Sigma$.}
\medskip \noindent In view of the boundary
condition (2.4) on the metric $q_{ab}$,
let us consider $\hat{q}_{ab} = r^\beta q_{ab}$ (and set $\hat{K}_{aA}{}^B =
K_{aA}{}^B$). Then, $\hat{q}_{ab} \sim  e_{ab} +O(1/r)$
(and $\hat{K}_{aA}{}^B \sim O(r^{-2+\beta/2})$.) Therefore, for the pair
$(\hat{q}_{ab}, \hat{K}_{aA}{}^B)$, we can apply arguments which are
completely parallel to the ones used in the 3+1-dimensional theory (see,
e.g. [26], section 3). The conclusion is the following: given a spinor field
$\lambda^o_{A}$ in the asymptotic region of $\Sigma$, which is constant
with respect to the connection ${D}^o$ defined by (an ${}^o\!E_{aA}{}^B$
compatible with) $e_{ab}$, there is a unique solution $\hat{\lambda}_A$
to (4.10) with respect to $(\hat{q}_{ab}, \hat{K}_{aA}{}^B)$ with the
asymptotic behavior $\hat{\lambda}_A \sim \lambda_A^o + O(1/r)$. Using the
conformal invariance of (4.10), we therefore conclude that there is a unique
solution $\lambda_A$ to (4.10) with respect to $(q_{ab}, K_{aA}{}^B)$,
with the asymptotic fall-off $\lambda_A \sim r^{\beta/4}
\lambda^o_A + O(1/r)$.

We can now return to (4.11) and the main argument. Let us suppose that
$\beta \le 0$. Then, {\it not only do the limits of all integrals in
(4.11) exist but the surface term goes to zero.} Therefore, for
$\beta\le 0$, the sum of the volume terms is zero. However, the first of
these terms is manifestly non-negative and our energy condition is precisely
that the integrand of the second term is also non-negative. Hence, each must
vanish in the limit $R_o\to \infty$. The vanishing of the first term
implies ${\cal D}_a\lambda_A =0$ everywhere on $\Sigma$. As in the
3+1-theory, (4.10) admits two solutions which are linearly independent almost
everywhere on $\Sigma$. The availability of two independent spinors which are
constant with respect to ${\cal D}$ implies that the curvature $F_{abA}{}^B$ of
${\cal D}$ must vanish, which in turn implies that the matter terms must vanish
and that the initial data is that for Minkowski space. Thus,  if $\beta \le 0$,
we must have $\beta = 0$: for $\beta < 0$, there are no (globally well-defined)
solutions to the constraints satisfying the asymptotic conditions if the matter
fields are to obey our energy condition. Thus, we have established the
desired result.

As we indicated above, the final argument is somewhat different from that
in the 3+1-dimensional theory. In particular, we do not have a manifestly
positive expression for energy in the case $\beta >0$; both the surface
and the volume integrals in (4.11) diverge in that case!
\bigskip\bigskip

{\bf 5. Conclusion}

In the last three sections, we analysed the Hamiltonian formulation of general
relativity coupled to matter in 2+1-dimensions in the asymptotically flat
context. The analysis we presented can also be carried out in the
connection-dynamics framework which is in fact simpler in 2+1 dimensions
since all canonical variables can be taken to be real [25]. Furthermore,
the proof of the positive energy theorem  would have been more direct in that
framework. However, since some of
the results are rather unexpected, we chose to
present the material in the
more familiar geometrodynamical language to emphasize
the fact that they are not artifacts of connection-dynamics.

We saw that the Hamiltonian framework differs from that in 3+1 dimensions in a
number of important respects. First, the boundary conditions on the geometrical
fields are such that,
while we can fix an Euclidean metric $e_{ab}$ near infinity,
if the mass in the space-time is non-zero, the
physical 2-metrics $q_{ab}$ do not
approach it even asymptotically. The two are related by a conformal factor
$r^{-\beta}$ which goes to
zero or diverges at infinity depending on the sign of
$\beta$. Therefore, the construction of the Hamiltonian framework is somewhat
more involved. In particular, the asymptotic symmetry group is just the
two-dimensional Abelian group of time translation and spatial rotation.

We carried out a detailed analysis of constraints of the theory and found that
their role is somewhat different from that in the 3+1 theory. The smeared
vector constraint  is differentiable on the phase space even when the
smearing  shift field $N^a $ remains asymptotically bounded, i.e., even when
$N^a\sim N^a_o(\theta) + O(1/r)$. Thus, the diffeomorphisms generated by all
such shifts $N^a$ are to be
regarded as gauge in the 2+1 theory. In the 3+1 case,
by constrast, only the diffeomorphisms generated by shifts which {\it vanish}
asymptotically that are regarded as gauge;
the generators of asymptotically constant
vector fields are the ADM 3-momenta. In the present case, there is no conserved
quantity analogous to the 3-momentum in agreement
with the fact that the asymptotic
symmetry group does not admit space translations. Spatial rotations, on
the other hand are not generated by constraints since their fall-off is given
by
$N^a \sim r$. They do induce canonical transformations on the phase space whose
generating function can be obtained by adding a surface term to the constraint
functional. The generator is the angular momentum.
Thus, when we are ``on shell,''
angular momentum is given by a surface integral at infinity.

In the case of the scalar constraint, we found that the differentiability
requirement forces the lapse to go to zero at infinity at a rate that depends
on $\beta$. Thus, to obtain a constraint function which is differentiable
on the entire phase space, the lapse has to vanish faster than any
inverse power of $r$. A more significant surprise is that, if we ask that
the lapse be asymptotically constant, say $N=1 +O(1/r)$ so that it
corresponds to an unit time translation at infinity, the resulting
infinitesimal motion defined formally on the phase space, although formally
defined, fails to be a canonical transformation unless $\beta <2$. Thus, the
Hamiltonian framework simply fails to exist if $\beta >2$. (In retrospect,
therefore, without loss of generality, we could have added the requirement
$\beta <2$ in the boundary condition (2.4), i.e., in the very construction
of the configuration sapce ${\cal C^{\rm geo}}$.) We then focussed on the
``physical part'' of the phase space where $\beta <2$ and computed the
Hamiltonian generating the unit time translation. We found that it can be
obtained by adding a surface term to the smeared constraint. The value of
the surface term is simply $\beta/8G$. Thus, on the physically relevant phase
space, when the constraints are satisfied,
the numerical value of the Hamiltonian
is bounded from above by $1/4G$. Finally, using $SU(1,1)$ spinors, we analyzed
the issue of the lower bound. Using an argument along the lines given by Witten
[17] in the 3+1-dimensional theory, we showed, if the matter fields satisfy a
local energy condition, $\beta$, and
hence the value of the Hamiltonian on physical
states, is necessarily non-negative,
vanishing if and only if space-time is globally
Minkowskian. Thus, on physical states,
the Hamiltonian is bounded by $0<H< 1/4G$.

We will now discuss the implications of these results to general relativity in
3+1-dimensions.

As we recalled in section 1, 3+1-dimensional vacuum general relativity
in presence of a space-like Killing field is equivalent to
2+1-dimensional general relativity coupled to certain scalar fields [11]
(which satisfy our energy condition). If the spatial Killing field in the
3+1-dimensional theory is translational, the induced geometry in 2+1-dimensions
can be expected to be asymptotically flat. An exhaustive analysis of such
space-times was carried out recently under the assumption that there is an
additional axial Killing field, and it was shown, in particular, that
there exists a large class of examples
in which the scalar fields have spatially
compact support [21]. In all these cases, we can use our expression of the
Hamiltonian to represent the energy per unit length
(along translational isometry)
in 3+1-dimensional gravity waves. In particular, contrary to what one might
have initially expected, this energy is bounded from above.

For concreteness and simplicity, let us restrict our detailed discussion to
cylindrical waves [4-6] where both $\partial/\partial z$ and $\partial/\partial
\theta$ Killing fields are hypersurface orthogonal.
In the 2+1-dimensional picture
these space-times correspond to gravity coupled to a single scalar field, where
the scalar field as well as the geometry
have an additional rotational symmetry.
In this case, one can go to coordinates
$t,r,\theta$ with $-\infty < t<\infty, \
0\le r<\infty,\  0\le \theta <2\pi$,
in which the space-time metric takes the form:
$$ ds^2 = e^{\Gamma (r,t)}(-dt^2 + dr^2) + r^2 d\theta^2 \ ,\eqno(5.1)$$
where the coefficent $\Gamma$ is completely determined by the
scalar field $\Phi$
via:
$$\Gamma (r,t) = \textstyle {1\over 2}\int_o^r dr' r'
 [(\partial_t \Phi)^2  + (\partial_{r'} \Phi)^2 ] ,\eqno(5.2)$$
the integral being performed on a $t={\rm const}$ slice.
The scalar field $\Phi$ satisfies
the wave equation with respect to the space-time
metric (5.1). However, as remarked in section 1, because
of axi-symmetry, this is
equivalent to the condition that it satisfy
the wave equation with respect to the
globally Minkowskian metric, obtained by setting $\Gamma (r,t) = 0$, i.e., with
respect to
$$dS_o^2 = -dt^2 +dr^2 +r^2 d\theta^2  \ .\eqno(5.3)$$
Because of this, the problem decouples: one can first solve for the scalar
field
$\Phi$ on the Minkowski space (5.3), construct $\Gamma$ from $\Phi$ using
(5.2),
and just write down the resulting metric (5.1)
to obtain a solution to the combined
Einstein-scalar field equations.
(A similar decoupling occurs also when the Killing
fields are not hypersurface orthogonal.)

Thorne's $c$-energy [7] is easy to express in this framework: apart from
an overall
constant, it is just the conserved
energy associated with $\Phi$ propagating on the flat metric (5.2)
$$c := \textstyle{1\over 16} \int_0^\infty r dr [(\partial_t \Phi)^2  +
(\partial_r \Phi)^2 \eqno(5.4)$$
which, however, is to be interpreted as the energy associated with the coupled
system, consisting of gravity and the scalar field. It is obvious that, the
$c$-energy is non-negative, vanishes if and only if $\Phi =0$ and $dS^2$ is the
flat Minkowskian metric $dS_o^2$, and that, even if one restricts oneself to
scalar fields with compact spatial support, it is {\it unbounded} from above.

Let us compare it with our Hamiltonian. For the metric (5.1), the value of the
Hamiltonian reduces to:
$$H = {1\over 4G}(1 - e^{- 4Gc}) \eqno(5.5)$$
Thus, the relation is non-polynomial!
However, $H$ {\it is} a monotonic function
of $c$; both attain the value zero --their minimum-- simultaneously and as $c$
tends to infinity $H$ tends to its upper bound $1/4G$. In the weak field limit,
where the field $\Phi$ and hence the $c$-energy can be taken to be small
compared to $1/G$, the two agree. However, as one scales up $\Phi$, space
``curls up'' and the ``gravitational contribution'' to the energy becomes
significant. The total energy then is quite different from the $c$-energy. Note
also that
{\it the boundedness of the Hamiltonian is a genuinely non-perturbative
result}. Indeed, if we expand out the exponent, we obtain a power series in
$G$:
$$ H = c -2Gc^2 + {8\over 3}G^2c^3+ .... ,\eqno(5.6) $$
where, the individual terms, being proportional to the powers of $c$, are all
unbounded. One can take the $c$-energy as a function on the (gauge fixed) phase
space and ask for the canonical transformation it generates. Since it is a
function only of $H$,
one would expect it also to correspond to a time-evolution
for {\it some} lapse. This expectation is correct. The lapse is simply $N =
\exp {\Gamma/2}$,
which tends asymptotically to $\exp 4c$. Thus, while the lapse
corresponding to the Hamiltonian $H$ is asymptotically identity (by the very
definition of the Hamiltonian), that corresponding to $c$ is not; even its
asymptotic value is a ``q-number'' ---it depends on the phase space variables.
Furthermore, as we approach the bound $\beta = 2$, the lapse corresponding to
the c-energy diverges. Within symplectic geometry, this is the origin of the
unboundedness of the c-energy. Note also that $H$ is defined more generally,
e.g., in the case when the 4-geometry
has only one (space-translational) Killing
field which is not necessarily hypersurface orthogonal.
Finally, the example of
cylindrical waves brings out the fact that although the Hamiltonian is bounded,
there is nothing unusual about time; it is not cyclic.
This is because the points
at which $H=0$ and $H= 1/4G$ are not identified; they correspond to entirely
different geometries.

Let us briefly compare our results with those obtained in the  twistorial
approach to ``quasi-local'' quantities [27]. Cylindrical waves have been
analysed by Tod [28] in this framework. He found that in the limit appropriate
to obtaining the total, ADM-like energy per unit length, the prescription of
[27] yields twice the $c$-energy and is thus non-polynomially related to our
Hamiltonian. This is
perhaps not surprising because it is known that the results
of [27] are not always in agreement with those obtained by Hamiltonian methods.
Quasi-local expressions which are geared to Hamiltonian methods were proposed
in [29]. It would be interesting to evaluate them for cylindrical waves and
compare the result with the one obtained here.

Finally, in this paper we have restricted ourselves to the behavior of the
2+1-dimen\-sion\-al gravitational field at spatial infinity. A similar analysis
can be carried out also at null infinity and again leads to some results which
are surprising from a 3+1-dimensional perspective [30].

\bigskip\bigskip

{\bf Acknowledgements} We would like to
thank Jiri Bi\v{c}\'ak, Piotr Cru\'sciel,
Karel Kucha\v r, Bernd Schmidt and Paul Tod for discussions. This work was
supported in part by the NSF grant PHY-93-96246 and by the Eberly research
funds of  Penn State University.

\bigskip
\bigskip

{\bf References}

\item{[1]}R. Beig and W. Simon, Commun. Math. Phys. {\bf 78}, 75 (1980);
P. Kundu, J. Math. Phys. {\bf 22}, 1236 (1981).
\item{[2]}A. Ashtekar, R. Tate and C. Uggla, Int. J. Mod. Phys. {\bf D2},
15 (1993).
\item{[3]}J. Barrow, Phys. Rep. {\bf 85}, 1 (1979); J. Pullin in {\it
Relativity and Gravitation: Classical and Quantum}, edited by J. D' Olivio
{\it et al}. (World Scientific, Singapore, 1991).
\item{[4]}K. S. Thorne, Phys. Rev. {\bf 138}, 251 (1965).
\item{[5]}J. Stachel, J. Mat. Phys. {\bf 7}, 1321 (1966).
\item{[6]}K. Kucha\v{r}, Phys. Rev. {\bf D4}, 955 (1971).
\item{[7]}R. Arnowitt, S. Deser and C. W. Misner in {\it Gravitation:An
Introduction to Current Research}, edited by L. Witten (Wiley, New York,
1962).
\item{[8]}H. H. Bondi, M. G. J. van der Burg and A. W. K. Metzner, Proc.
Roy. Soc. (London) A{\bf 269}, 21 (1962).
\item{[9]}R. Penrose, Proc. Roy. Soc. (London), A{\bf 284}, 159 (1965).
\item{[10]}B. Berger, Ann. Phys. (N.Y.) {\bf 83}, 458 (1974);
B. Berger, Ann. Phys. (N.Y.) {\bf 156}, 155 (1984).
\item{[11]}D. Kramer, H. Stephani, E. Herlt and M. MacCallum in {\it Exact
Solutions of Einstein's field equations}, (Cambridge U.P., Cambridge,
1980), p 326.
\item{[12]}M. Allen, Class. Quantum Grav. {\bf 4}, 149 (1987).
\item{[13]}G. Bonacina, A. Gamba and M. Martellini, Phys. Rev. {\bf D45},
3577 (1992).
\item{[14]}S. Deser, R. Jackiw and G. 't Hooft, Ann. Phys. (N.Y.) {\bf 152},
220 (1984).
\item{[15]}C. W. Misner, Private Communication at the Brill-Misner Symposium
(1993).
\item{[16]}J. Wheeler in:{\it Relativity,, Groups and Topology}, edited by
C. DeWitt and B. DeWitt (Blackie \& sons, London, 1964), Section on Brill
waves.
\item{[17]}E. Witten, Commun. Math. Phys. {\bf 80}, 381 (1981).
\item{[18]}M. Henneaux, Phys. Rev. D{\bf 29}, 2766 (1984).
\item{[19]}T. Regge, C. Teitelboim, Ann. Phys. (NY) {\bf 88}, 276 (1974);
A. Ashtekar,
{\it New Perspectives in Canonical Gravity} (Bibliopolis, Naples, 1988).
\item{[20]}R. Jantzen, Nuovo Cimento {\bf 55} 161 (1980).
\item{[21]}B. Berger, P. Cruis\'cel, V. Moncrief and H. J. Seifert,
gr-qc/9404005
\item{[22]}P. M. A. Dirac, {\it Lecture notes on quantum mechanics},
(Yashiva Univewrsity Press, New York 1964).
\item{[23]}A. Comet and G. Gibbons, Nucl. Phys. B{\bf 299}, 719 (1988).
\item{[24]}R. Penrose and W. Rindler, {\it Spinors and Spacve-time},
(Cambridge University Press, Cambridge, 198 )
\item{[25]}A. Ashtekar, {\it Non-perturbative Canonical Gravity} (World
Scientific, Singapore, 1991).
\item{[26]}A. Ashtekar and G. T. Horowitz, J. Math. Phys. {\bf 25}, 1473
(1984).
\item{[27]}R. Penrose, Proc. Roy. Soc. (London) A{\bf 381}, 53 (1982).
\item{[28]}K. P. Tod, Class. Quan. Grav. {\bf 7}, 2237 (1990).
\item{[29]}A. J. Dougan and L. J. Mason, Phys. Rev. Lett. {\bf 67}
2119 (1991).
\item{[30]}A. Ashtekar, J. Bi\v{c}\'ak and B. G. Schmidt (in preparation).

\end